# Exploring Fermi Surface Nesting and the Nature of Heavy Quasiparticles in the Spin-Triplet Superconductor Candidate CeRh$_2$As$_2$

Bo Chen,[1] Hao Liu,[1] Qi-Yi Wu,[1] Chen Zhang,[1] Xue-Qing Ye,[1] Yin-Zou Zhao,[1] Jiao-Jiao Song,[1] Xin-Yi Tian,[1] Ba-Lei Tan,[1] Zheng-Tai Liu,[2] Mao Ye,[2] Zhen-Hua Chen,[2] Yao-Bo Huang,[2] Da-Wei Shen,[3] Ya-Hua Yuan,[1] Jun He,[1] Yu-Xia Duan,[1] and Jian-Qiao Meng[1,*]

[1]*School of Physics, Central South University, Changsha 410083, Hunan, China*
[2]*Shanghai Synchrotron Radiation Facility, Shanghai Advanced Research Institute, Chinese Academy of Sciences, 201204 Shanghai, China*
[3]*National Synchrotron Radiation Laboratory and School of Nuclear Science and Technology, University of Science and Technology of China, Hefei 230026, China*
(Dated: Friday 15th March, 2024)

In this study, we investigate the electronic structure of a spin-triplet superconductor candidate CeRh$_2$As$_2$ using high-resolution angle-resolved photoemission spectroscopy and density functional theory calculations. Notably, Fermi surface nesting hints at connections to magnetic excitation or quadrupole density wave phenomena, elucidating the superconducting mechanisms. Measured band structures reveal primarily localized 4$f$ electrons, with minor itinerant contributions. Additionally, a transition from localized to itinerant behavior and significant $c$-$f$ hybridization anisotropy underscore the role of $f$-electrons in shaping electronic properties. These findings deepen our understanding of CeRh$_2$As$_2$'s unconventional superconductivity and magnetism. Further exploration promises advances in superconductivity research.

CeRh$_2$As$_2$, a recently discovered heavy-fermion (HF) superconductor, has attracted considerable attention due to its exceptional characteristics [1]. With decreasing temperature, CeRh$_2$As$_2$ undergoes a potential quadrupole density wave (QDW) phase transition at $T_0 \approx$ 0.4-0.5 K [2, 3], succeeded by a superconducting transition at $T_c \approx$ 0.3-0.4 K [1–6], and finally an antiferromagnetic (AFM) phase transition at the Néel temperature $T_N \approx$ 0.25 K, coexisting with superconductivity within its superconducting phase [4, 5]. Multiple field-induced superconducting phases have been observed, with the transition between two different superconducting phases suggested to be a spin-singlet (even-parity) to spin-triplet (odd-parity) transition [1]. Despite its remarkably low superconducting transition temperature, CeRh$_2$As$_2$ displays an unusually high superconducting upper critical field, reaching up to 14 T [1], surpassing the Pauli limit. This multiphase superconductivity in HF compounds is rare, as most unconventional superconductors typically exhibit a single superconducting phase [7, 8]. Moreover, in systems where superconductivity and magnetism coexist or compete, the superconducting phase generally lacks AFM order parameters, with $T_N$ consistently exceeding the superconducting temperature $T_c$.

The distinct physical properties of CeRh$_2$As$_2$, uncovered through numerous experimental [1–6] and theoretical [9–17] studies, stem from its unique local non-centrosymmetric crystal structure. Unlike other non-centrosymmetric compounds like CeRhSi$_3$ [18], CePt$_3$Si [19], and CeCoGe$_3$ [20], which showcase a combination of odd-parity and even-parity states without displaying multiphase superconductivity, CeRh$_2$As$_2$ exhibits multiphase superconductivity. This distinction renders CeRh$_2$As$_2$ an excellent model system for investigating how crystal structures lacking local central symmetry can influence unconventional superconductivity and magnetic interactions, thus elucidating the complex interplay between these phenomena [4]. For instance, the locally symmetry-lacking structure of CeRh$_2$As$_2$ may induce a significant Rashba-like spin-orbit coupling (SOC) of the 4$f$ and the conduction bands, a factor believed to be pivotal in the superconductivity parity transition [1, 10, 11].

Despite numerous experimental [1–6] and theoretical [9–17] efforts, the mechanism of superconductivity in CeRh$_2$As$_2$ remains controversial. The observation of multiphase superconductivity in the spin-triplet candidate 5$f$-HF UTe$_2$ [21–23] underscores the pivotal role of spin fluctuations in its unconventional superconducting state. CeRh$_2$As$_2$'s crystal structure suggests that spin fluctuations may play a key role in its superconducting state [1]. AFM order picture has been proposed to explain the transition between its two superconducting states [17]. Neutron scattering has revealed the existence of quasi-two-dimensional (2D) AFM spin fluctuations with a wavevector of $q = (\pi/a, \pi/a)$ [24]. Some studies suggest a close relationship between the Kondo effect and superconductivity in CeRh$_2$As$_2$ [16], contributing to the ongoing debate over the role of Kondo physics [2, 10–15]. Specific heat measurement suggested that the heavy $f$ electrons are involved in the superconducting transition [1]. The resistivity shows a characteristic hump at $\sim$ 40 K, and the magnetic entropy monotonically increases to reach the value $R\ln4$ at 60 K, suggesting Kondo screening is active well above $T_N$ and involving the two low-lying doublets with very close energy in the crystal electric field (CEF) state [2]. Angle-resolved photoemission spectroscopy (ARPES), a momentum-resolved technology, serves as an ideal tool to address these issues, identifying potential Fermi surface (FS) nesting required for magnetic excitation [24] and QDW, detecting band splitting caused by spin-orbit coupling (SOC), and elucidating the role played by $f$-electrons.

In this paper, we investigate the electronic structure of CeRh$_2$As$_2$, a spin-triplet superconductor candidate, using high-resolution ARPES measurements and DFT calculations.

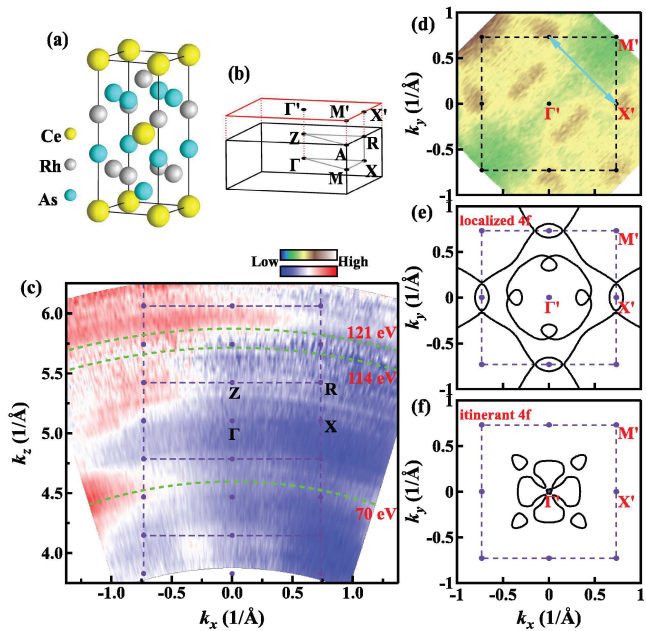

FIG. 1. FSs of CeRh$_2$As$_2$ at low temperature of 11 K. (a) Crystal structure of CeRh$_2$As$_2$. (b) 3D BZ (black) and projected surface BZ (red) of CeRh$_2$As$_2$, with marked high-symmetry momentum points (black dots). (c) Experimental 3D FS maps measured using $h\nu$ = 48-130 eV photons in 2 eV steps, in the $\Gamma ZRX$ plane. Final state arcs (green-dashed lines) for different photon energies are indicated. (d) Constant photon energy $k_x$-$k_y$ map taken with 70 eV photons. Photoemission intensities were integrated over an [-10 meV, 10 meV] energy window with respect to $E_F$. The cyan double arrow indicate the possible FS nesting vector $q$. (e) and (f) Calculation 2D FS contours at $k_z$ = 0.5 $\pi/c$ [25], considering 4$f$ electrons as core electrons (localized 4$f$) and valence electrons (itinerant 4$f$), respectively.

We observe a FS nesting vector that aligns with magnetic excitation, potentially associated with QDW phenomena. Furthermore, the observed consistency between the measured FS topology and localized 4$f$ electrons calculations, along with the involvement of $f$-electrons in shaping the FS as evidenced by on-resonance measurements, collectively suggest the dual nature of the 4$f$ electrons.

In this study, high-quality single crystals of CeRh$_2$As$_2$ were grown via a Bi-flux method. ARPES measurements were conducted at the BL09U and BL03U beamlines of the Shanghai Synchrotron Radiation Facility (SSRF) using a Scienta DA30 analyzer under ultra-high vacuum conditions ($< 4\times10^{-11}$ mbar). All samples were cleaved *in situ* at $\sim$ 11 K. FS topology was mapped by systematic measurements with varying photon energies ($h\nu$ = 48-130 eV) for $k_z$ (perpendicular) and constant photon energy ($h\nu$ = 70 eV) for $k_\parallel$ (horizontal) directions. Experimental FS topology and band structures were compared with density-functional theory (DFT) calculations. Temperature-dependent on-resonance 4$d \to$ 4$f$ ($h\nu$ = 121 eV) ARPES spectra were applied to probe the nature of Ce 4$f$ electrons.

CeRh$_2$As$_2$ crystallizes in a centrosymmetric CaBe$_2$Ge$_2$-type tetragonal structure with the $P4/nmm$ space group (No.129, D74h) [Fig. 1(a)] [30]. The Ce site lacks local inversion symmetry, while maintaining a global spatial inversion center. Positioned between two different RhAs block layers, this local inversion symmetry breaking at the Ce layer is considered crucial for CeRh$_2$As$_2$'s superconductivity. The bulk Brillouin zone (BZ) of CeRh$_2$As$_2$ and its projection onto the (001) surface BZ, including high-symmetry momentum points, are shown in Fig. 1(b). At 11 K, photon-energy-dependent normal scans was illustrated in Fig. 1(c). Although the intensity of the FSs varies with the photon energies, it behaves in three dimensions as predicted by calculation (See Supplemental Material for more data [25]). Additionally, like other studies [31–33], the FSs near the BZ center exhibit quasi-2D characteristics. Subsequently, an inner potential ($V_0$) of 16 eV was taken, consistent with findings from other studies [31–33], a common value in Ce-based HFs.

To search for the possible spin fluctuation required FS nesting vector $q = (\pi/a, \pi/a)$, a constant photon energy $k_x$-$k_y$ mapping was conducted using 70 eV photons at 11 K [Fig. 1(d)]. The high-intensity spots at the $X'$ correspond to a nesting vector identical to the wave vector $q$ required by AFM spin fluctuations [24], as indicated by the cyan double arrow, which may also account for the QDW [2]. This observation aligns with findings from other ARPES studies [32]. Figures 1(e) and 1(f) depict the calculated FS at $k_z$ = 0.5 $\pi/c$, considering 4$f$ electrons as core electrons (localized 4$f$) and valence electrons (itinerant 4$f$), respectively. The experimental FSs is inferred to be more consistent with the localized 4$f$ DFT calculation. This finding reinforces the justification for treating 4$f$ electrons as core electrons in non-resonant excitation studies. The FS shape calculated based on localized 4$f$ electrons also supports the possibility of FS nesting with a wave vector $q$ [Fig. 1(e)].

To investigate the properties of Ce 4$f$ electrons, we conducted 4$d \to$ 4$f$ on-resonant ARPES measurements using 121 eV photons for resonance enhancement. Figures 2(a) and 2(b) compare off-resonant (114 eV) and on-resonance (121 eV) photoemission spectra at 11 K, along the $\Gamma'$-$X'$ direction. In the off-resonance spectrum, contributions from Rh 4$d$ and As 4$p$ states create dispersive bands with non-$f$ orbital character, corroborated by the angle-integrated energy distribution curve (EDC) (solid blue line) in Fig. 2(d). To facilitate comparison with DFT calculations, in Fig. 2(a), we have superimposed the experimental band structure with dispersions along $\Gamma'$-$X'$ direction calculated for localized 4$f$ electrons (white-hashed lines), as the position measured by 114 eV photons corresponds to $k_z \sim$ 0 [Fig. 1(c)]. While the calculated electronic structure generally aligns with the ARPES results, there are still some notable discrepancies. In particular, the $\Gamma'$ centered electron-like dispersion (cyan-dashed line) indicated by the red arrow does not correspond to the theoretically calculated energy band. Considering its energy scale and momentum range, it appears to align well with the energy band along the $X$-$M$ direction (see Fig. S1 of the Supplemental Material [25]). The pronounced nesting feature, consistent with the magnetic excitation $q$ revealed by Chen et al [24], may also be responsible for the QDW [2].

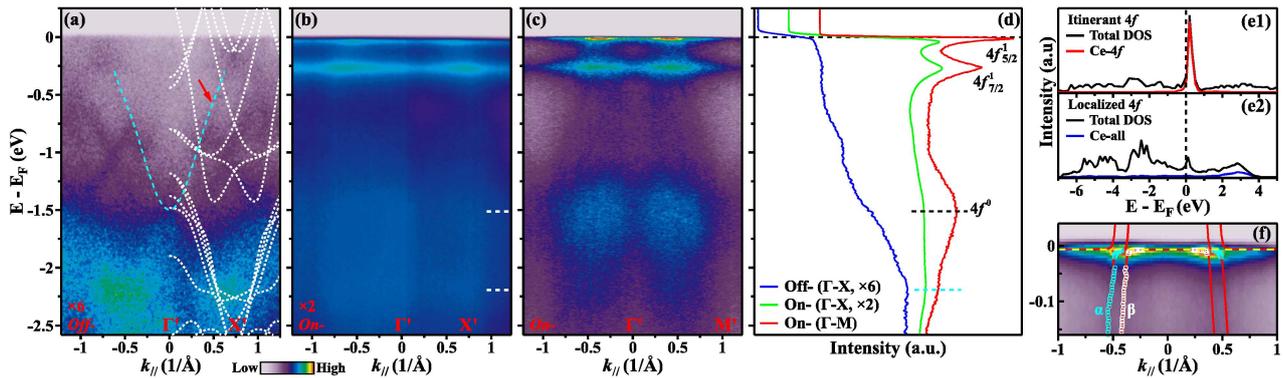

FIG. 2. Off-resonance and on-resonance ARPES data of CeRh$_2$As$_2$ at 11 K. (a) Off-resonance (114 eV) ARPES data along the $\Gamma'$-$X'$ direction, with overlaid localized $4f$ calculation results along the $\Gamma$-$X$ direction. (b) and (c) On-resonance (121 eV) ARPES data along the $\Gamma'$-$X'$ and $\Gamma'$-$M'$ directions, respectively. (d) Angle-integrated EDCs for data in (a)-(c). Positions of the $4f^1_{5/2}$, $4f^1_{7/2}$, and $4f^0$ states are indicated. (e1) and (e2) Calculated density of state (DOS) vs energy $E$ for CeRh$_2$As$_2$, utilizing itinerant and localized $4f$ calculation methods, respectively. (f) Photoemission intensity near the $E_F$. The $c$-$f$ hybridization is modeled using the periodic Anderson model. Squares and open circles represent positions of the conduction and hybridized $f$ bands, respectively. The yellow dashed line indicates the position of the $f$-level.

The on-resonance data exhibit a significant enhancement of Ce $4f$ states, attributed to strengthened Ce $4f$-electron photoconduction matrix elements. Observations reveal two strong heavy quasiparticle bands arising from the splitting of $4f^1$ final state induced by SOC, as confirmed by the angle-integrated EDC (solid green line) in Fig. 2(d). A state approximately 0.27 eV below $E_F$ is assigned to the $4f^1_{7/2}$ state, while the one proximite to $E_F$ corresponds to the $4f^1_{5/2}$ state. Similar observations have been reported in previous studies on CeRh$_2$As$_2$ [33? ], and analogous structures have been documented in other Ce-based HF systems [34–42].

In Figure 2(b), short dashed lines indicate the emergence of two broader peaks approximately 1.5 eV and 2.2 eV below $E_F$. The peak around 1.5 eV below $E_F$ corresponds to the $f^0$ state, resulting from pure charge excitations of the trivalent Ce ion ($4f^1 \rightarrow 4f^0$) [34–42]. Meanwhile, the peak around 2.2 eV below $E_F$ represents a shoulder of the $f^0$ state, reflecting hybridization spreading due to the structure in the valence band density of states [Fig. 2(a)] [34, 35]. Remarkably, the intensity of the $f^1$ state surpasses that of the $f^0$ state, differing from observed patterns in CePt$_2$In$_7$ [36, 37], CeRhIn$_5$ [36, 38], Ce$_2M$In$_8$ ($M$ = Co, Rh, and Ir) [36, 39], and others. This departure, consistent with observations in CeCoIn$_5$ [36, 42] and CeRh$_6$Ge$_4$ [41], implies increased itinerancy of $4f$ electrons in CeRh$_2$As$_2$, according to the periodic Anderson model (PAM). However, DFT calculations [Fig. 2(e)] suggest that the majority of Ce $4f$ states in CeRh$_2$As$_2$ are situated above $E_F$, with only a small fraction below $E_F$. This suggests a limited degree of $4f$ electron itinerancy, a common feature in Ce-based HFs. Compared to Yb-based [43, 44] and U-based [44–47] HFs, the $f$ electrons in CeRh$_2$As$_2$ are more localized.

In Figures 2(b) and 2(c), the significantly momentum-dependent intensity of the $4f^1_{5/2}$ and $4f^1_{7/2}$ states indicates pronounced $c$-$f$ hybridization, suggesting the involvement of $f$ electrons in forming the Fermi surface. Along the $\Gamma'$-$X'$ direction, the $4f^1_{5/2}$ state exhibits weaker strength compared to the $4f^1_{7/2}$ state, while measurements along the $\Gamma'$-$M'$ direction [Fig. 2(c)] reveal a contrasting trend. This disparity suggests anisotropic hybridization strength in momentum space, like due to the local non-centrosymmetric crystal structure of CeRh$_2$As$_2$.

Figure 2(f) presents the on-resonance spectral data along the $\Gamma'$-$M'$ direction near $E_F$, highlighting on the intersection between the conduction bands and the $f$ band (See Fig. S5 in Supplemental Material [25]). The data is normalized by the corresponding resolution-convoluted Fermi-Dirac function (RC-FDD). The unique observation of hybridization between the two adjacent conduction bands ($\alpha$ and $\beta$) and the $f$ band is evident. Consistent with prior investigations [35, 48], we analyzed the on-resonance spectra using a hybridization band model derived from the periodic Anderson model (PAM) [48, 49]. In this framework, the hybridization results in ($E^+$) and lower ($E^-$) bands, expressed as

$$E^{\pm} = \frac{\varepsilon_f + \varepsilon_k \pm \sqrt{(\varepsilon_f - \varepsilon_k)^2 + 4|V_k|^2}}{2},$$

where, $\varepsilon_f$ denotes the renormalized $f$-level energy ($4f^1_{5/2}$ Kondo resonance state), $\varepsilon_k$ represents the bare conduction band, and $V_k$ symbolizes the renormalized hybridization (half of the direct hybridization gap) [48]. Fitting the model yields $\varepsilon_f$ = -9 meV for both the $\alpha$ and $\beta$ bands, with $V_k$ values of 19 ± 3 meV and 8.5 ± 3 meV for the $\alpha$ and $\beta$ bands, respectively. The obtained hybridization parameter value for the $\alpha$ band is consistent with values reported in a prior study ($\sim$ 20 meV) [31] and is similar to values found in other Ce-based heavy fermion compounds [35, 42, 50].

Understanding the properties $f$-electrons in HF compounds and how they evolve with temperature is crucial for comprehending HF physics, especially in compounds CeRh$_2$As$_2$. Therefore, conducting a thorough investigation into $c$-$f$ hybridization in CeRh$_2$As$_2$ is of paramount significance. Figure 3(a) presents the temperature-dependent on-resonance ARPES measurement along the $\Gamma'$-$X'$ direction ranging from 11 K to 160 K. Throughout this temperature range, the $4f^1_{5/2}$ and $4f^1_{7/2}$ states are observable. The intensity of the $4f^1_{5/2}$ states decreases with increasing temperature, while the intensity of the $4f^1_{7/2}$ states shows little change. However, even at 160 K, their intensity still exhibits significant momentum dependence, indicating ongoing $c$-$f$ hybridization. Remarkably, the characteristics of the flat bands persist even at room temperature (See Fig. S4 in Supplemental Material [25]), suggesting that the formation of the heavy quasiparticle band initiates at a temperature significantly higher than its Kondo temperature ($T_K \sim 30$ K) [1, 2, 6], consistent with observations in other Ce-based [38–42], Yb-based [43], and U-based [46, 47] HF systems.

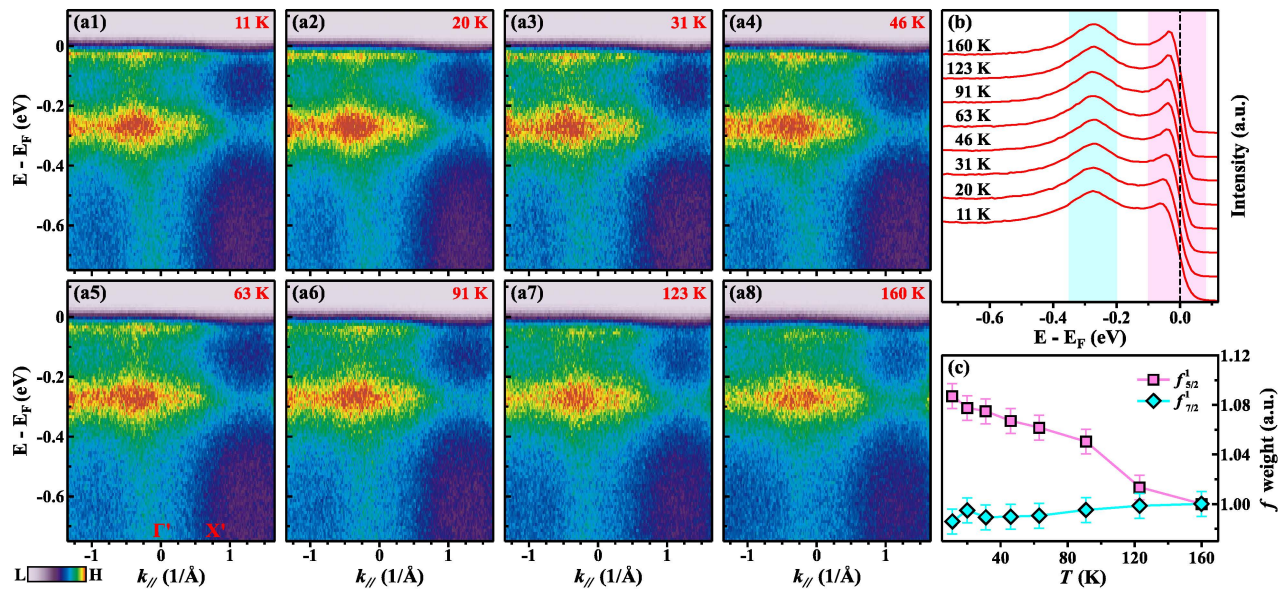

FIG. 3. Temperature-dependent evolution of heavy quasiparticle bands in CeRh$_2$As$_2$. (a) On-resonant band structure along the $\Gamma'$-$X'$ direction at specified temperatures. (b) Angle-integrated EDCs over the angle range delineated in (a) across different temperatures. (c) Normalized quasiparticle spectral weight as function of temperature, normalized to the reference value at 160 K. Cyan and magenta markings signify spectral weight integrals corresponding to the highlighted regions in (b).

Figure 3(b) illustrates the temperature evolution of the angle-integrated EDCs within the measured momentum range shown in Fig. 3(a). As the temperature decreases, the peak of the $4f^1_{7/2}$ state remains nearly unchanged, while the peak of the $4f^1_{5/2}$ state becomes sharper. This observation suggests a broad temperature range crossover corresponding to the localized-to-itinerant transition of $4f$ electrons. While the enhanced hybridization temperature can be attributed to Kondo screening involving the excited crystal electric field (CEF) state, no visible serrated peak features caused by CEF splitting are observed in the EDCs, likely due to the limitations in our measurement's energy resolution.

Figure 3(c) provides a quantitative analysis of the evolution of $f$-electron spectral weight with temperature, normalized to the highest temperature data. Within the experimental accuracy, the spectral weight of the $4f^1_{7/2}$ state shows no significant change with temperature. Conversely, the spectral weight of the $4f^1_{5/2}$ state gradually increases as temperature decreases, reaching nearly 10% in the displayed temperature range. These temperature-evolution behaviors are considerably weaker than those observed along the $\Gamma'$-$M'$ direction [31], emphasizing the pronounced anisotropy of c-f hybridization. The observed slight increase may originate from its interaction with an electron-type conduction band along the $\Gamma'$-$X'$ direction [Fig. 2(a)], where the band bottom closely approaches the $4f$ band. This band crossing and hybridization could lead to the emergence of additional hybridized state(s) slightly above $E_F$, contributing to the integrated intensity at higher temperatures and giving the impression of reduced temperature dependence of the $f^1_{5/2}$ state.

In summary, the investigation into the electronic structure of CeRh$_2$As$_2$, a spin-triplet superconductor, utilizing ARPES and DFT calculations, has yielded crucial insights into its multiphase superconductivity and antiferromagnetism. A significant observation is the presence of Fermi surface nesting, suggesting a potential link to magnetic excitation or quadrupole density wave phenomena, thus illuminating the mechanisms driving the superconducting state. The measured band structures reveal the dual nature of $4f$ electrons, primarily localized with a minor itinerant contribution. Moreover, the broad transition from localized to itinerant behavior and the signif-icant anisotropy in $c$-$f$ hybridization underscore the importance of $f$-electrons in shaping the electronic properties of $CeRh_2As_2$. These findings highlight the intricate interplay among electronic states in $CeRh_2As_2$, advancing our understanding of its unconventional superconductivity and magnetism.

This work was supported by the National Natural Science Foundation of China (Grant No. 12074436), the National Key Research and Development Program of China (Grant No. 2022YFA1604204), the Science and Technology Innovation Program of Hunan Province (2022RC3068), and the Changsha Natural Science Foundation (Grant No. kq2208254). We are grateful for resources from the High Performance Computing Center of Central South University.

---

# Supplemental Materials:

**Exploring Fermi Surface Nesting and the Nature of Heavy Quasiparticles in the Spin-Triplet Superconductor Candidate CeRh$_2$As$_2$**

Bo Chen,[1] Hao Liu,[1] Qi-Yi Wu,[1] Chen Zhang,[1] Xue-Qing Ye,[1] Yin-Zou Zhao,[1] Jiao-Jiao Song,[1] Xin-Yi Tian,[1] Ba-Lei Tan,[1] Zheng-Tai Liu,[2] Mao Ye,[2] Zhen-Hua Chen,[2] Yao-Bo Huang,[2] Da-Wei Shen,[3] Ya-Hua Yuan,[1] Jun He,[1] Yu-Xia Duan,[1] and Jian-Qiao Meng[1,*]

[1]*School of Physics, Central South University, Changsha 410083, Hunan, China*
[2]*Shanghai Synchrotron Radiation Facility, Shanghai Advanced Research Institute, Chinese Academy of Sciences, 201204 Shanghai, China*
[3]*National Synchrotron Radiation Laboratory and School of Nuclear Science and Technology, University of Science and Technology of China, Hefei 230026, China*
E-Mail: jqmeng@csu.edu.cn

The supplemental materials to 'Angle-resolved photoemission spectroscopy study on the heavy quasiparticles in the locally non-centrosymmetric heavy fermion superconductor CeRh$_2$As$_2$' contains 'Single crystal synthesis', 'Band Structure Calculation', 'Comparison of the calculated dispersion', 'Comparison of the calculated Fermi surface contours', 'Measured $k_y$-$k_x$ constant energy contours', 'On-resonance ARPES spectral along Γ'-X' direction at high temperature', and 'On-resonance ARPES spectral along Γ''-M' direction'.

1. **Single crystal synthesis**

   High-quality single crystals of CeRh$_2$As$_2$ were grown synthesized using the Bi-flux method [1]. A mixture of high-purity elemental metals with a molar ratio of Ce(99.9%) : Rh(99.9%) : As(99.99%) : Bi(99.99%) = 1: 2 : 2 : 30 was loaded into an alumina crucible and sealed in a quartz tube. The ampule was heated to 1150 °C in a furnace and held at this temperature for 40 hours before being slowly cooled down to 700 °C at a rate of 1.5 °C/h. Upon reaching 700 °C, the assembly was removed from the furnace and promptly subjected to high-speed centrifugation to remove excess Bi flux. The resulting CeRh$_2$As$_2$ crystals exhibited a shiny, sheet-like morphology with typical dimensions of 1 × 1 × 0.2 mm$^3$.

2. **Band Structure Calculation**

   The calculations were performed using density functional theory (DFT) and implemented in the Vienna Ab initio Simulation Package (VASP) code [2,3]. Projector augmented-wave (PAW) potentials were employed within the generalized gradient approximation (GGA) in the Perdew, Burke, and Ernzerhof (PBE) parametrization [4]. A 16 × 16 × 8 *k*-point grid generated with the Monkhorst–Pack scheme was used for the summation in reciprocal space. The energy cutoff for the plane-wave expansion was set to 450 eV. To incorporate correlation effects of the Ce 4*f* electrons, the GGA+Hubbard-U (GGA + U) method was employed with U = 4 eV, a value commonly used in studies of Ce compounds. The Ce 4*f* electrons were treated both as core (localized) and valence (itinerant) electrons to extract the orbital contribution of 4*f* electrons. Paramagnetic order was assumed as the initial magnetic order, and spin-orbit interaction was considered in all calculations.

## 3. Comparison of the calculated dispersion

Figure S1 compares itinerant and localized DFT calculation results. In the itinerant 4*f* calculation (red solid lines), several renormalized flat bands appear slightly above $E_F$, indicating a substantial contribution from Ce-4*f* electrons. On the contrary, there are no *f*-band features within the energy range displayed in the localized 4*f* calculation (blue-dashed lines). For the occupied states, the itinerant 4*f* calculation is essentially identical to that of the localized 4*f* calculation, but bands in the itinerant 4*f* calculation exhibit a downward shift. We also observed that the presence of Ce 4*f* states near $E_F$ causes some modifications to the band structure very close to the $E_F$.

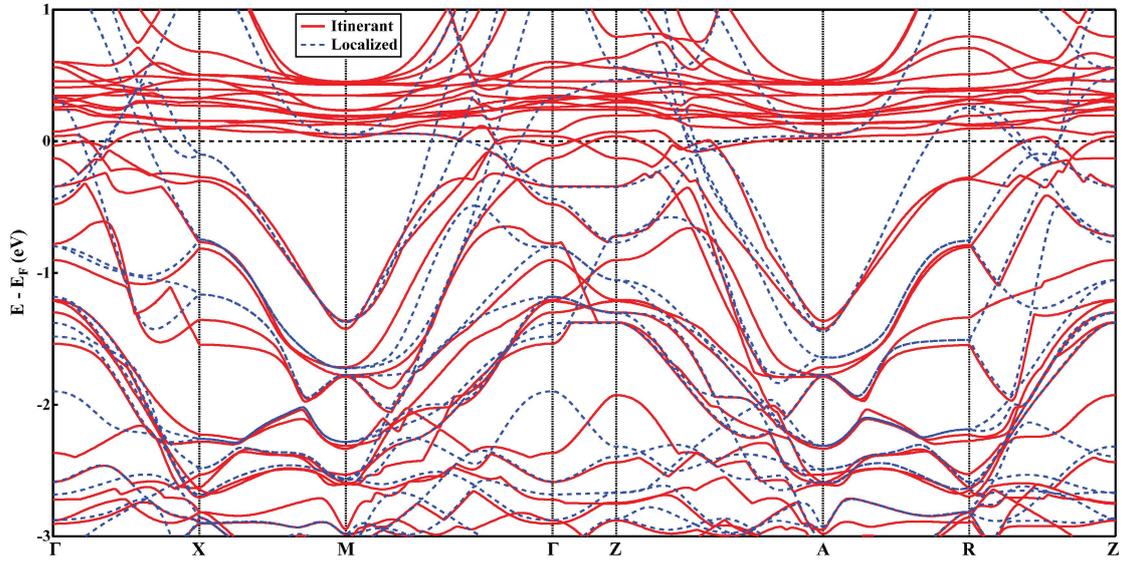

**Fig. S1** Comparison of itinerant and localized DFT calculation results. (a) and (b) band structure of CeRh$_2$As$_2$, considering 4*f* electrons as valence electrons (itinerant) and core electrons (localized), respectively.

## 4. Comparison of the calculated Fermi surface contours

Figure S2 displays the calculated FS of CeRh$_2$As$_2$ at $k_z = 0$ and $\pi/c$. The overall shape of the FSs exhibits a significant difference between the itinerant [Fig. S2(a)] and localized [Fig. S2(b)] DFT calculations. As depicted in Figs. S2(a1) and (a2), the $\Gamma$(Z)-centered small FSs exhibit quasi-2D characteristics, whereas the larger FSs near the BZ boundary display distinct three-dimensional features. The local 4$f$ calculated FSs show extensive parallel regions, which is likely to facilitate FS nesting, especially there are large parallel regions connectable by a wave vector identical to the magnetic excitation wave vector $q = (\pi/c, \pi/c)$.

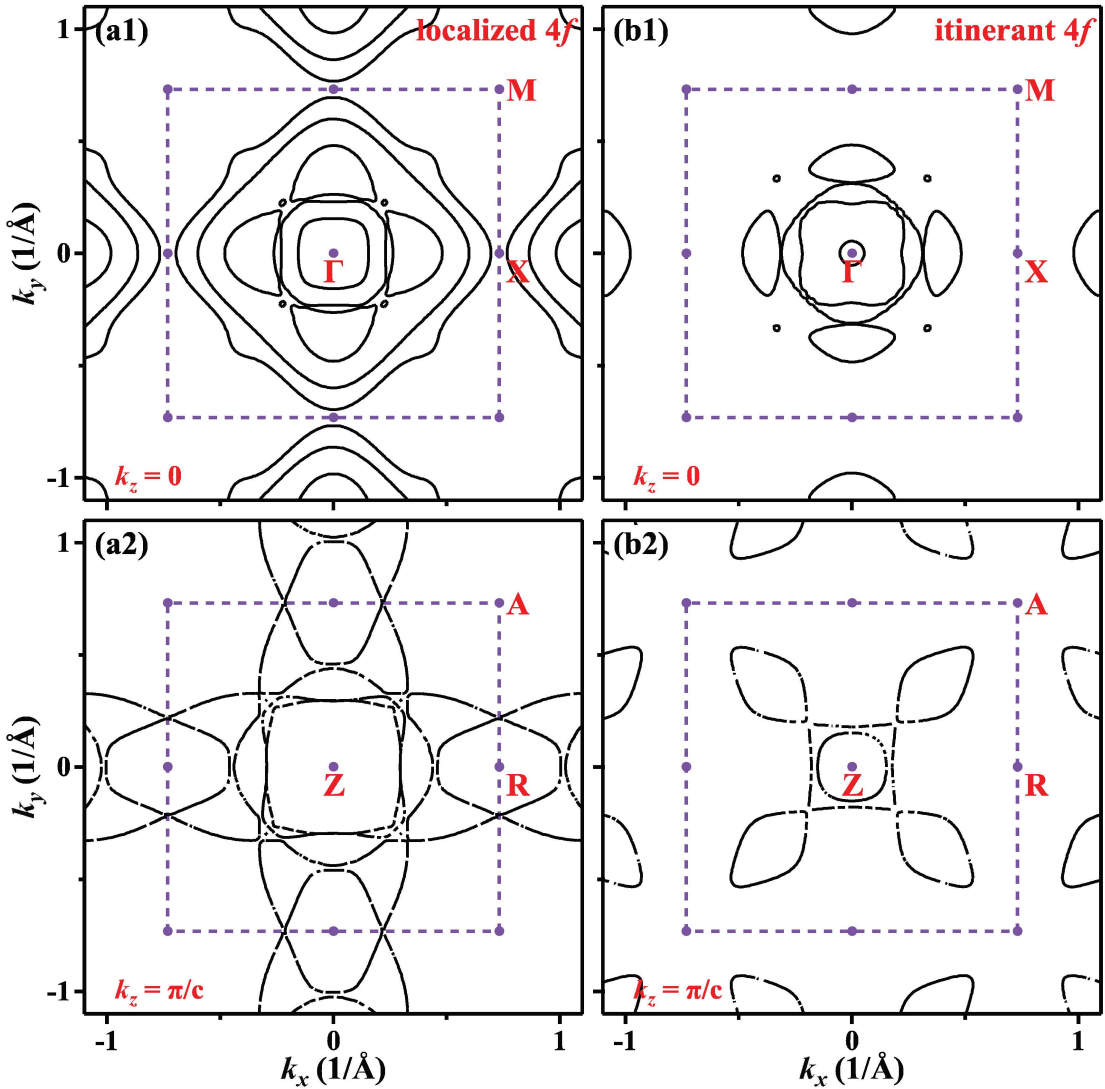

**Fig. S2** (a) and (b) Calculated 2D FS contours, considering 4$f$ electrons as core electrons (localized 4$f$) and valence electrons (itinerant 4$f$), respectively.

## 5. Measured $k_y$-$k_x$ constant energy contours

Figure S3 display the $k_x$-$k_y$ constant energy contours taken with 70 eV photons at a temperature of 11 K. All photoemission intensity data were integrated over an [−10 meV, 10 meV] energy window with respect to the marked energy. With increasing binding energy, the electronic pocket centered around M' becomes more prominent, exhibiting a progressively circular shape and diminishing size. As energy approaches $E_F$, the electron pockets centered at M' overlap with each other near X'.

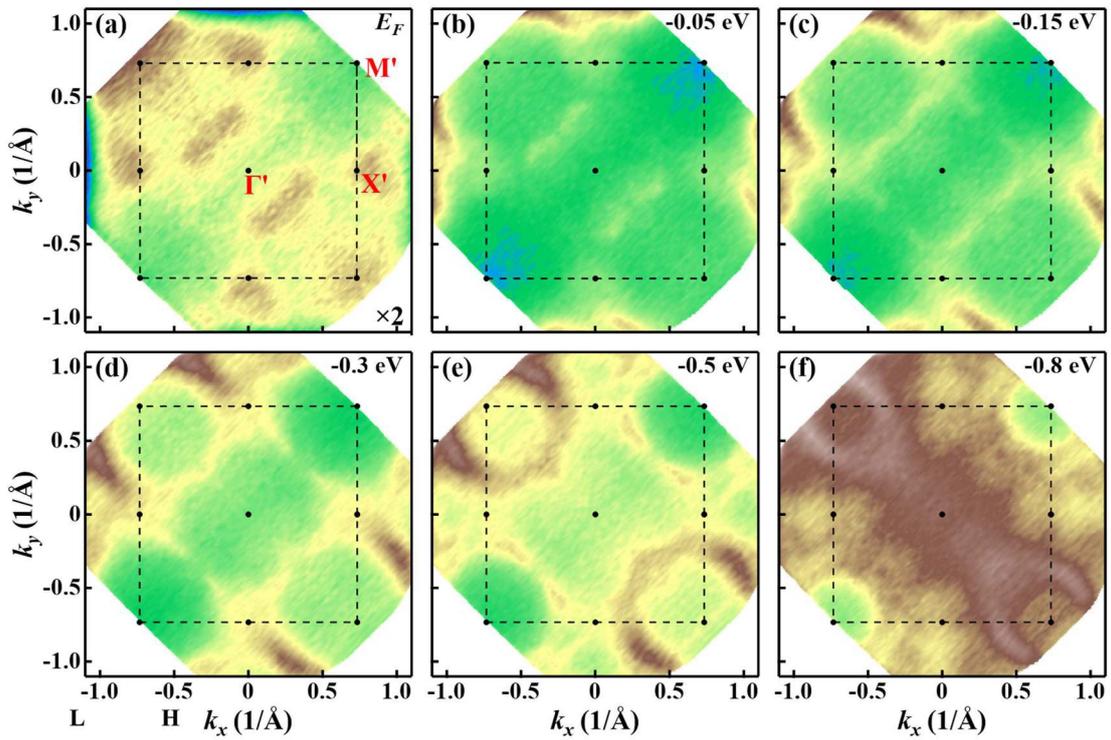

**Fig. S3** $k_x$-$k_y$ constant energy maps at marked energy obtained using 70 eV photons.

## 6. On-resonance ARPES spectral along Γ'-X' direction at high temperatures

Figure S4 display the heavy quasiparticle band structure along the Γ'-X' direction at high temperatures. Due to signs of sample aging at high temperatures, the high-temperature data was not utilized for quantitative analysis of the evolution of *f*-electron spectral weight with temperature. However, the two flat bands caused by spin-orbit splitting of the $4f^1$ final state remain present at high temperatures. Even at room temperature, as further evidenced by the integrated EDC in Fig. S4(c), the peaks corresponding to the $4f^1_{5/2}$ and $4f^1_{7/2}$ final states can still can be identified.

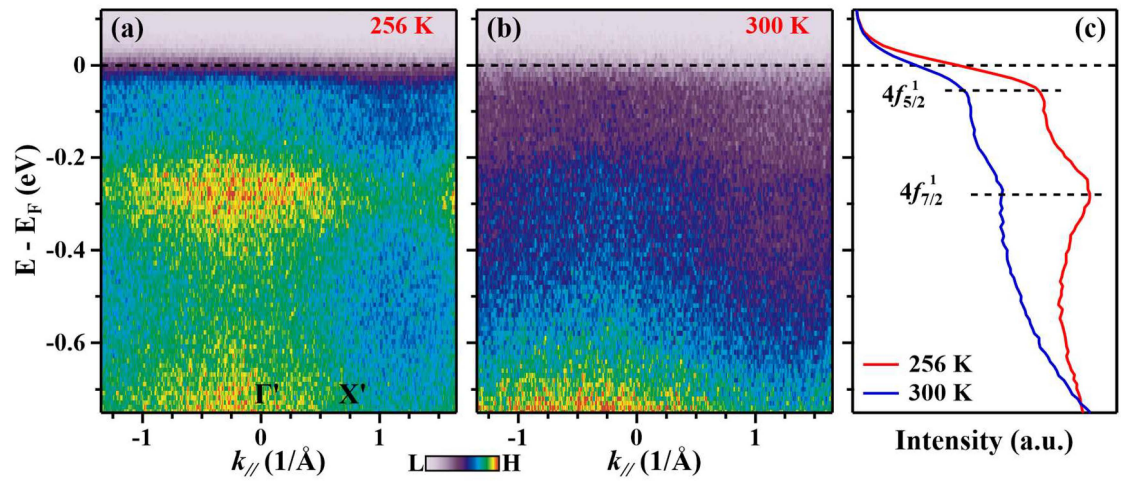

**Fig. S4** On-resonance ARPES data along the Γ'-X' direction at high temperatures of 256 K (a) and 300 K (b). (c) Angle-integrated EDCs over the shown momentum range. Positions of the $4f^1_{5/2}$ and $4f^1_{7/2}$ states are indicated.

## 7. On-resonance ARPES spectral along Γ'-M' direction at high temperatures

Figure S5 displays the heavy quasiparticle band structure along the Γ'-M' direction. In addition to the two clearer conduction bands on the outer side, there are also very weak conduction bands on the inner side closer to the Γ" point. Theoretical calculations indicate that the two outer conduction bands are caused by spin-orbit splitting of the same energy band.

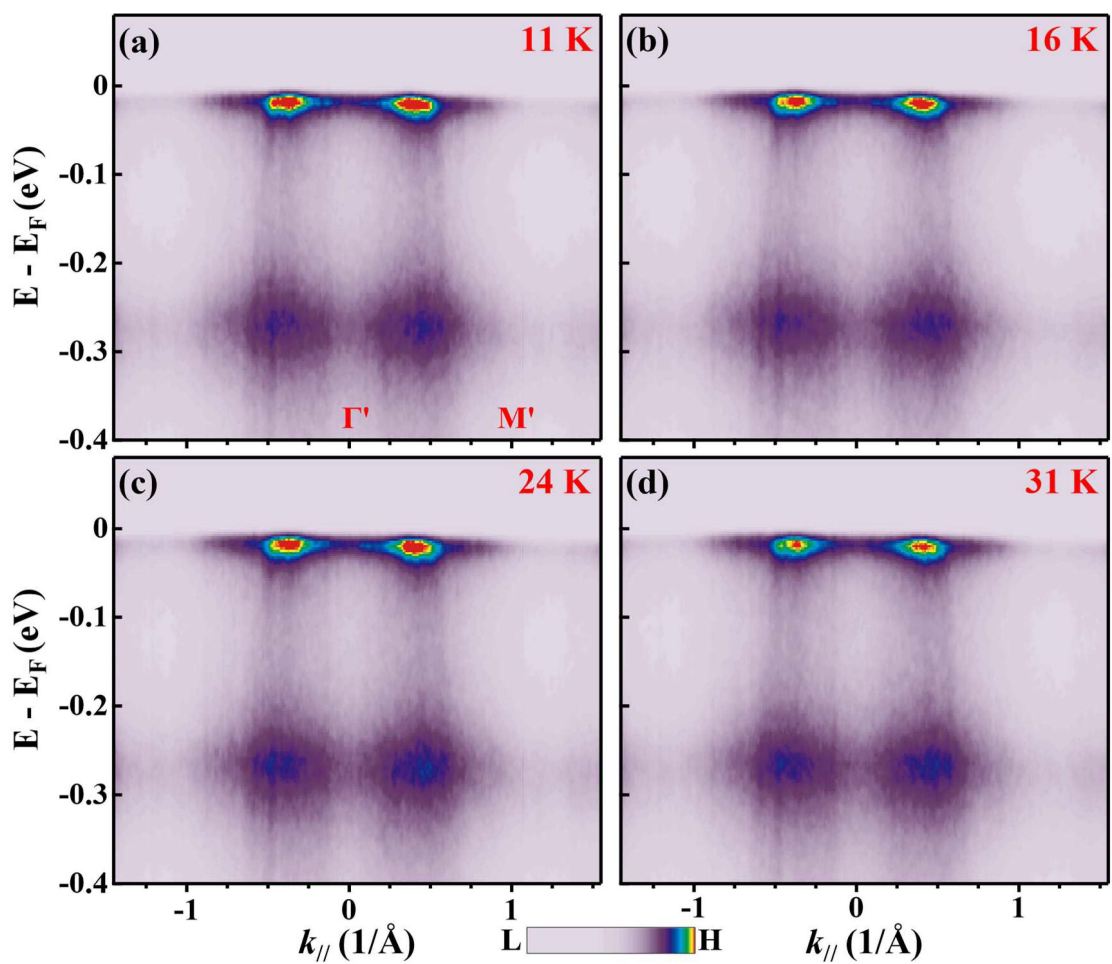

**Fig. S5** On-resonance ARPES data along the Γ"-M' direction at 11 K (a), 16 K (b), 24 K (c) and 31 K(d).